\documentclass{elsart}
\usepackage{epsfig} 

\newcommand{\be}{\begin{equation}}
\newcommand{\ee}{\end{equation}}
\newcommand{\ba}{\begin{eqnarray}}
\newcommand{\ea}{\end{eqnarray}}

\def\fun#1#2{\lower3.6pt\vbox{\baselineskip0pt\lineskip.9pt
  \ialign{$\mathsurround=0pt#1\hfil##\hfil$\crcr#2\crcr\sim\crcr}}}

\def\cal{}

\begin{document} 
\begin{frontmatter}
\title{Dark Energy Search with Supernovae} 
\author{Yun Wang} 
\address{Department of Physics \& Astronomy, Univ. of Oklahoma, 
440 W.~Brooks St., Norman, OK 73019, USA; wang@nhn.ou.edu} 

\begin{abstract} 

To determine the nature of dark energy from observational data, 
it is important that we use model-independent and optimal methods. 
We should probe dark energy using its density (allowed to
be a free function of cosmic time) instead of its equation of state.
We should minimize gravitational lensing effect on supernovae 
by flux-averaging. We need to include complementary data (for example,
from the Cosmic Microwave Background [CMB] and large scale structure [LSS])
in a consistent manner to help break the degeneracy between
the dark energy density and the matter density fraction. 
We should push for ambitious future supernova
surveys that can observe a large number of supernovae at
the highest possible redshifts.
I discuss these and other issues that will be important
in our quest to unravel the mystery of the nature of dark energy.

Current supernova, CMB, and LSS data already rule out dark energy
models with dark energy densities that vary greatly with time;
with the cosmological constant model providing an excellent fit
to the data. A precise measurement of dark energy density
as a free function of cosmic time will have a fundamental 
impact on particle physics and cosmology.

\end{abstract} 

\end{frontmatter}

\section{Introduction} \label{sec.intro}

The data of type Ia supernovae (SNe Ia) reveal that the expansion of our universe 
is accelerating \cite{Riess98,Perl99}.
This implies the existence of dark energy.  
Solving the mystery of the nature of dark energy will have revolutionary
implications in particle physics and cosmology.

The first most fundamental question about dark energy is whether
the dark energy density depends on time. 
The time dependence of the dark energy density can be used to 
distinguish the various models. \cite{Freese}  

A powerful probe of dark energy is SNe Ia, which can be calibrated
to be good cosmological standard candles 
\cite{Phillips,Riess95,Branch98}, and used
to measure how distance changes with redshift.
For a flat universe, the luminosity distance is 
\ba
\label{eq:dL(z)}
d_L(z) &=& cH_0^{-1} (1+z) \int_0^{z} \frac{dz'}{E(z')},\nonumber\\
E(z) &= &\left[ \Omega_m (1+z)^3 + (1-\Omega_m) \rho_X(z)/\rho_X(0)\right]^{1/2},
\ea
where $\rho_X(z)$ is the dark energy density. Note that dark nergy
only appears in the form of the function $E(z)$;
$\rho_X(z)$ is on the same footing as the matter density fraction
$\Omega_m$ in the context of observational tests.
The dark energy equation of state is \cite{Wang01a}
\be
\label{eq:wXz}
w_X(z)= \frac{p_X(z)}{\rho_X(z)}= \frac{(1+z)}{3} \frac{\rho_X'(z)}
{\rho_X(z)}-1.
\ee

\section{Probing Dark Energy Using Its Density Instead of Its Equation of State} 

Due to the integrals and parameter degeneracies in the distance-redshift relations 
of standard candles (see Eq.[\ref{eq:dL(z)}]), it is very difficult to detect the 
time-dependence of the dark energy equation of state $w_X(z)$ in a 
model-independent manner from realistic data \cite{Maor01,Barger}. 
By measuring the dark energy density $\rho_X(z)$, instead of $w_X(z)$, 
we effectively remove one integral, and increase the likelihood
to detect the evolution of dark energy. \cite{Wang01a,Tegmark02}
This would allow us to determine whether dark energy is vacuum energy, 
or some other form of exotic energy. 

It is easier to extract $\rho_X(z)$ from the data than to extract  
$w_X(z)$. This is easy to see mathematically: 
to obtain $\rho_X(z)$, take a single derivative of data (see Eq.[\ref{eq:dL(z)}]); 
to obtain $w_X(z)$, we need to take a second derivative 
(see Eq.[\ref{eq:wXz}]).\footnote{Note 
that in our analysis, no actual derivatives of data
are taken; the best fit model is found by using a likelihood
analysis based on Markov Chain Monte Carlo (MCMC).
However, the mathematical relation between 
$\rho_X(z)$ or $w_X(z)$ and the observable $d_L(z)$ (through
the apparent peak luminosity of SNe Ia) explains why 
$\rho_X(z)$ should be better constrained than $w_X(z)$.}

Fig.1 explicitly demonstrates the advantage of the $\rho_X(z)$ parametrization
over the $w_X(z)$ parametrization, using recent observational data.
Fig.1 shows $\rho_X(z)$ and $w_X(z)$ measured from
192 SNe Ia from the Tonry/Barris sample \cite{Tonry,Barris}, 
flux-averaged with $\Delta z=0.05$, and combined with CMB (shift parameter 
$R_0 = 1.716 \pm 0.062$ \cite{Spergel})
and LSS data (growth parameter $f_0 \equiv 
f(z=0.15) =0.51\pm 0.11$ \cite{Hawkins,Verde}).
The solid and dashed lines enclose the 68.3\% and 95\% confidence
regions respectively.\cite{WangMukherjee,WangFreese}
\begin{figure}[!hbt]
\begin{center} 
\psfig{file=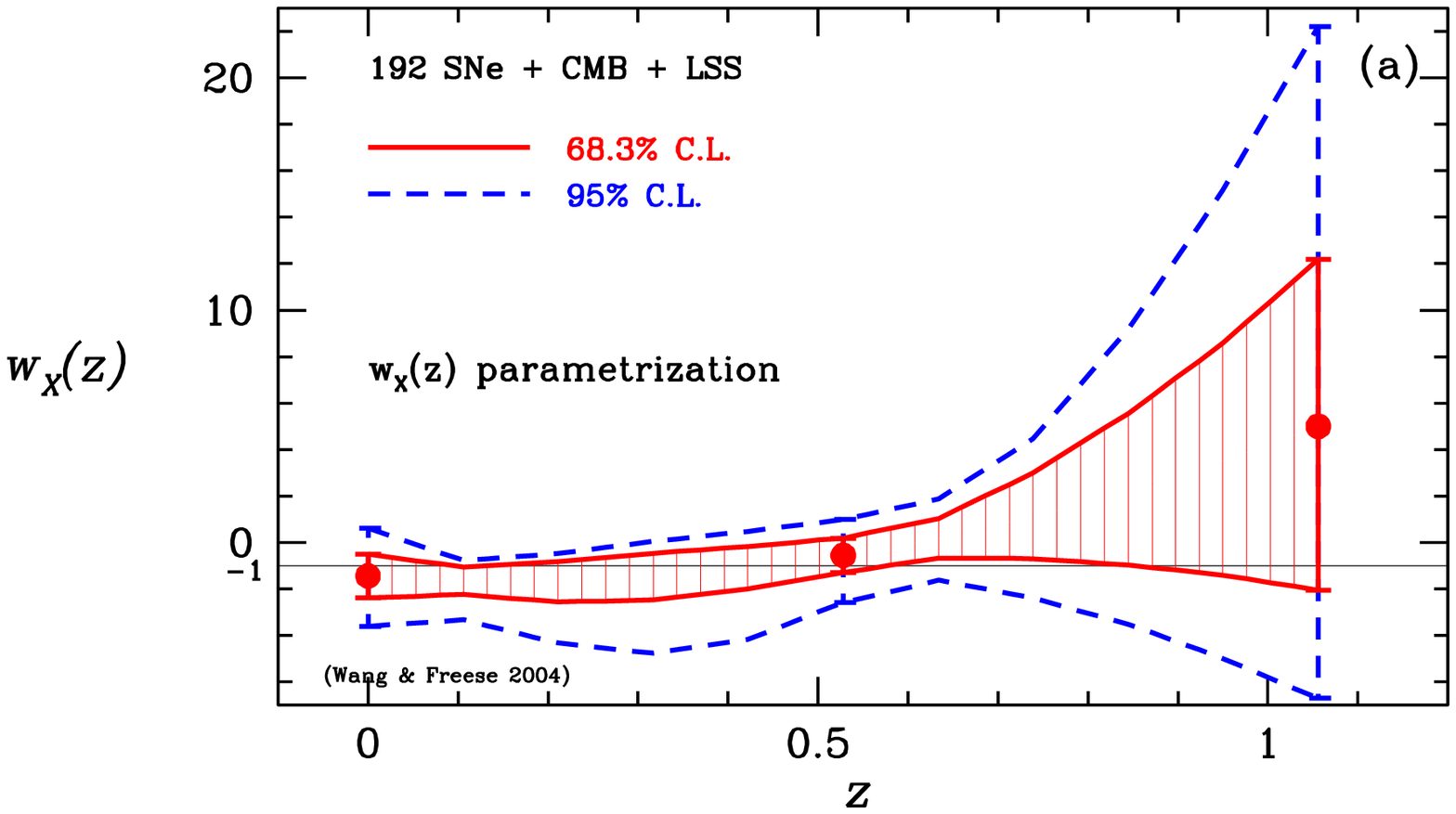,width=6.8cm} 
\psfig{file=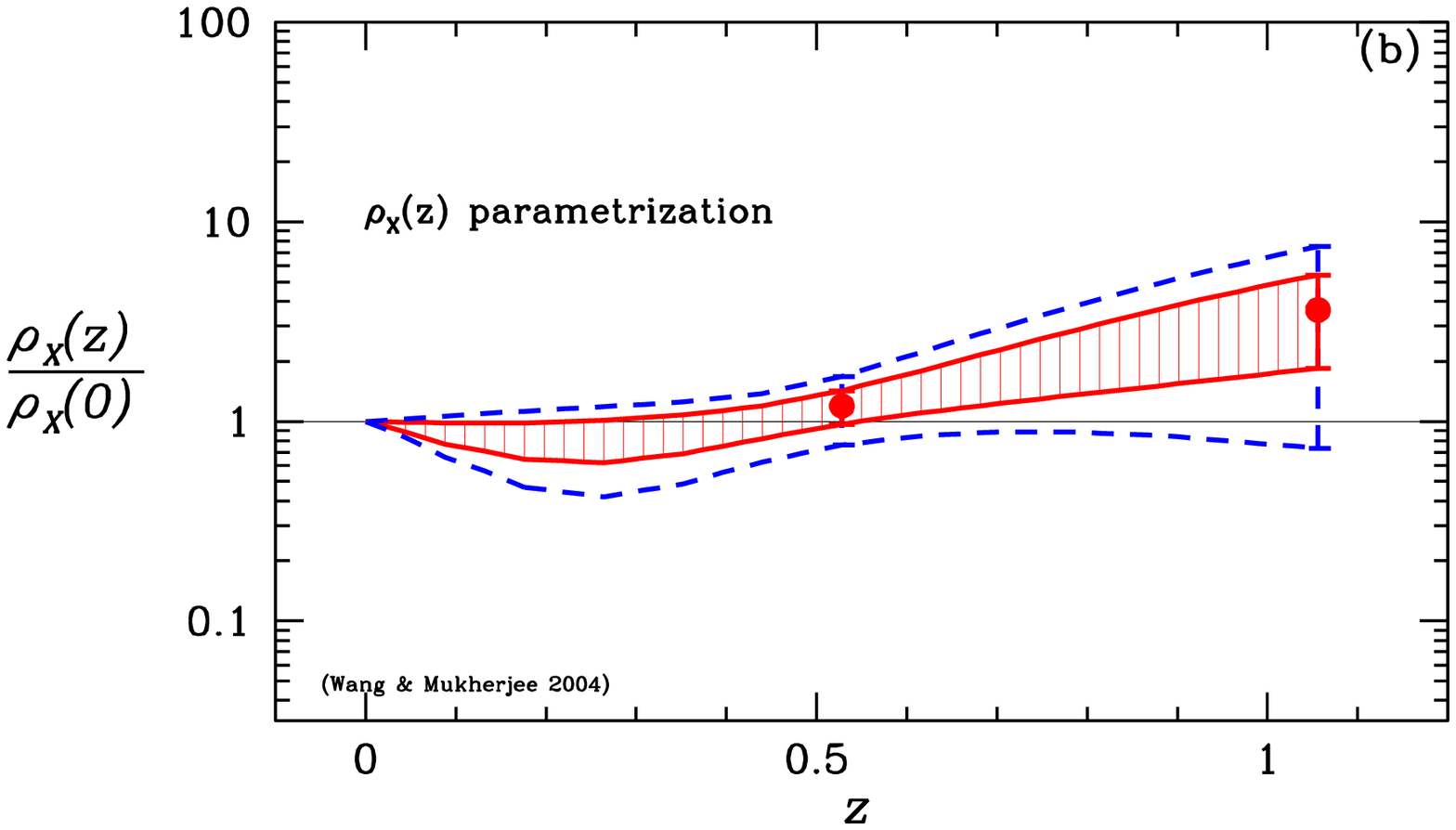,width=6.8cm} 
\vskip -3cm
\caption{\footnotesize%
$w_X(z)$ and $\rho_X(z)$ measured from
192 SNe Ia from the Tonry/Barris sample, 
flux-averaged, and combined with CMB (shift parameter 
$R_0 = 1.716 \pm 0.062$)
and LSS (growth parameter $f_0 \equiv 
f(z=0.15) =0.51\pm 0.11$) data.
} 
\end{center} 
\end{figure}
Note that $\rho_X(z)$ is more tightly constrained than $w_X(z)$.
\cite{Daly} and \cite{Huter} found similar results using a different sample of
SNe Ia (the Riess sample).

While using $\rho_X(z)$ in the likelihood analysis of data
encompasses {\it all} possible dark energy models,
using $w_X(z)$ implies $\rho_X(z)\geq 0$ 
since $\rho_X(0) > 0$ (see Eq.[\ref{eq:wXz}]).
To allow for exotic origins of dark energy,
we should not impose a positive prior on
the dark energy density.
This again argues in favor of using $\rho_X(z)$
instead of $w_X(z)$ to probe dark energy.\cite{WangTegmark}

\section{Go Deep: the Optimal SN Observational Strategy 
for Probing Dark Energy Evolution}

To determine whether SNe Ia are good cosmological standard candles, 
we need to nail the systematic uncertainties (luminosity evolution, 
gravitational lensing, dust) \cite{Aldering}.
This will require at least hundreds of SNe Ia at $z>1$. 
This can be easily accomplished by doing a supernova pencil beam survey 
using a dedicated telescope, which can be used for other things simultaneously 
(weak lensing, gamma ray burst afterglows, etc) \cite{Wang00a}. 

The inclusion of a large number of SNe Ia at high redshifts is
important for probing the time evolution of dark energy density.
Fig.2 shows that a shallow and wide survey (the survey strategy
for SNAP as of 2001) compares poorly with a supernova pencil beam survey
in constraining dark energy density at high $z$ \cite{Wang01b}.

\begin{figure}[!hbt]
\begin{center} 
\psfig{file=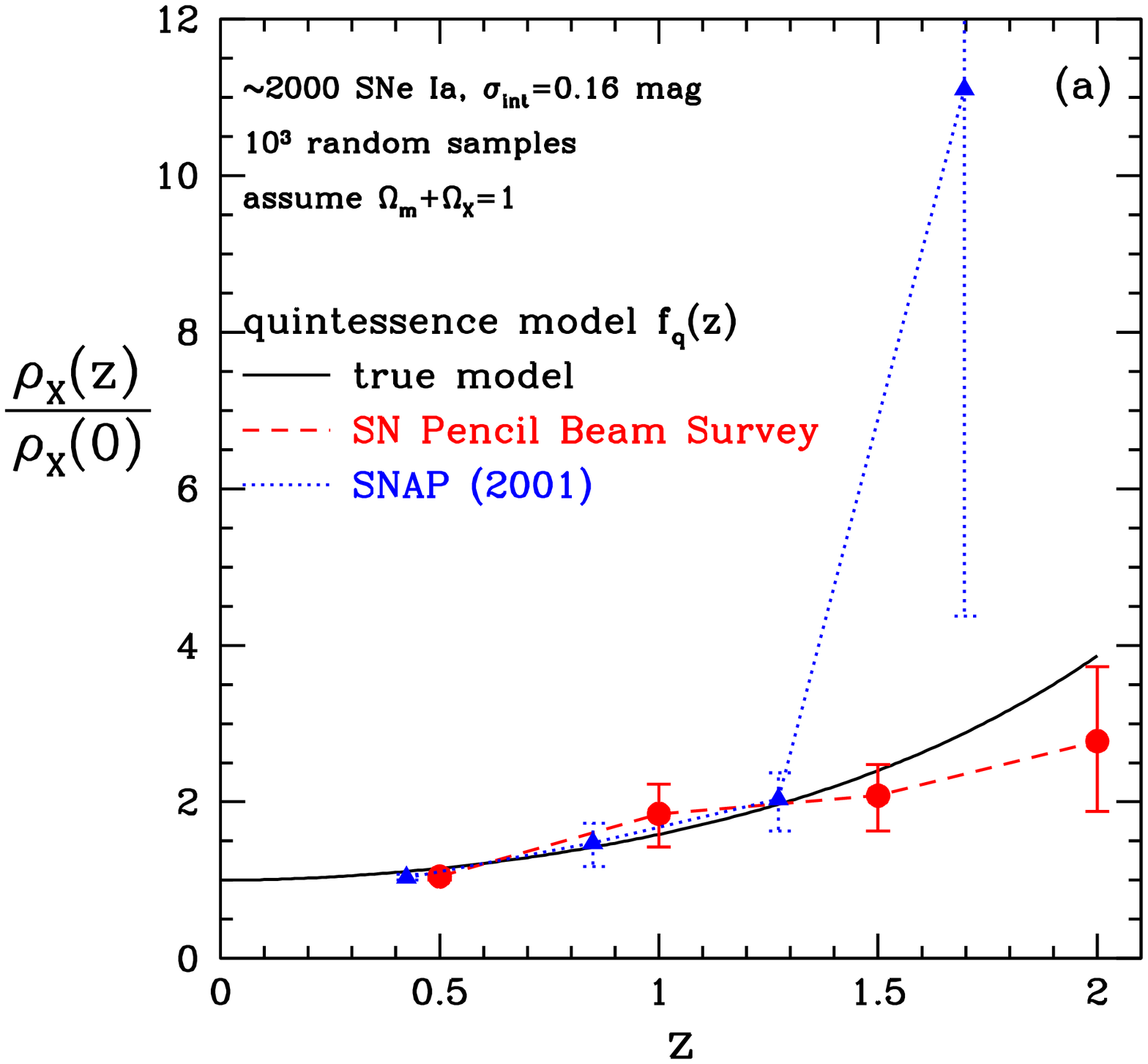,width=6.8cm} 
\psfig{file=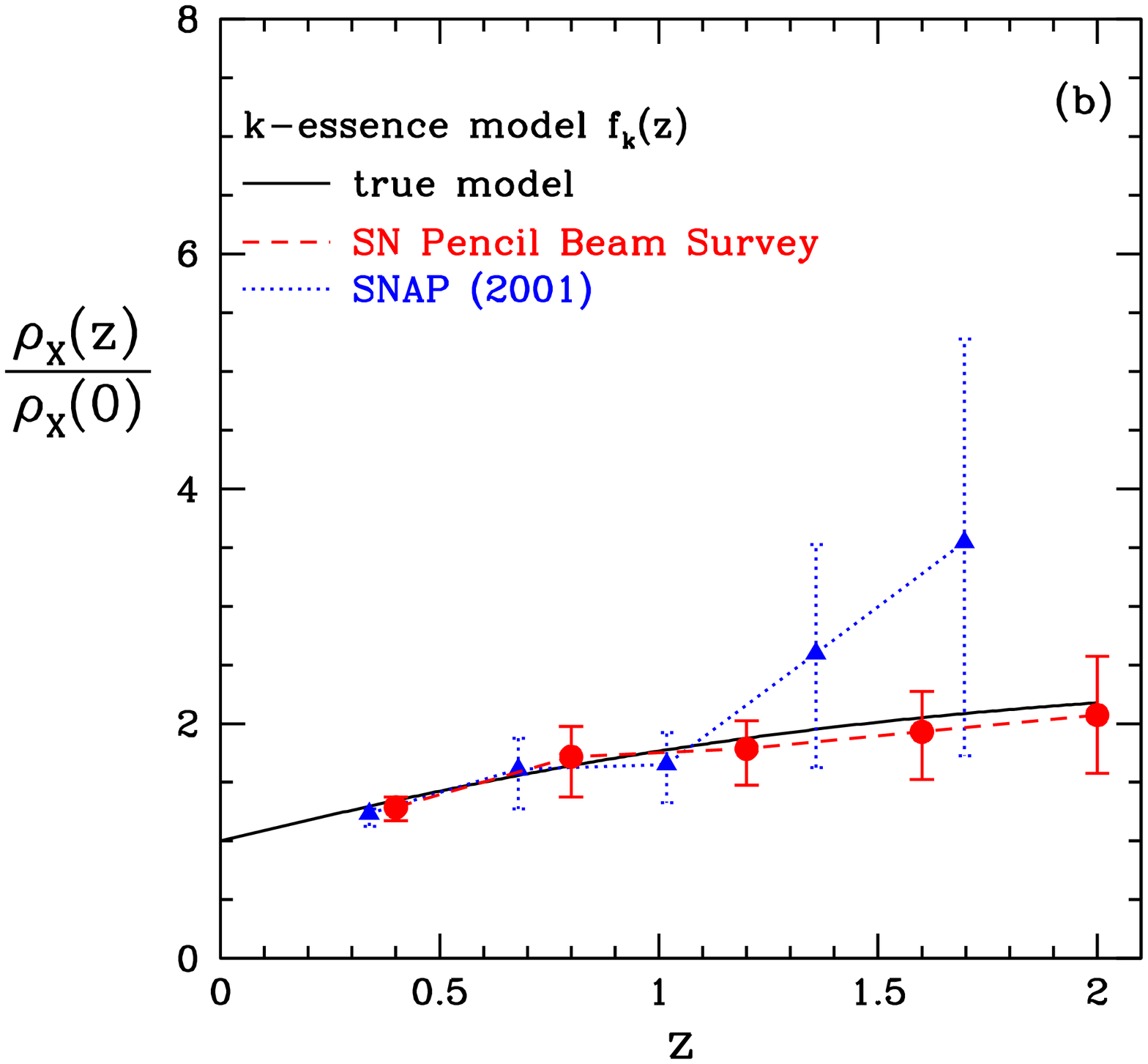,width=6.8cm} 
\vskip -0.6cm
\caption{\footnotesize%
A shallow and wide survey (the original survey strategy
for SNAP) compares poorly with a supernova pencil beam survey
in constraining dark energy density at high $z$ \cite{Wang01b}.
} 
\end{center} 
\end{figure}

Clearly, in order to probe time evolution of dark energy density,
it is essential to conduct a SN survey that is as deep as
feasible, so that the largest possible number of SNe Ia at
high redshifts can be observed.\footnote{The current SNAP survey 
strategy has been revised to image
all the SNe Ia upto a redshift of $1.7$ \cite{Tarle},
similar to a supernova pencil beam survey \cite{Wang00a,Wang01b}.}

\section{Data Analysis Method}

Here is a recipe for optimally extracting dark energy density $\rho_X(z)$ from 
observational data \cite{WangMukherjee}:\\
\noindent
 1) Obtain SN Ia peak brightnesses with uncertainties in 
 flux.\footnote{Flux averaging is only straightforward if the intrinsic 
 distribution of SN Ia peak brightness is Gaussian in flux (consistent
 with current data) \cite{WangMukherjee}.} \\
\noindent 
 2) Parametrize $\rho_X(z)$ as a free function (given by $\rho_X(z_i)/\rho_X(0)$,
 $i=1,2,...,N$, and interpolated elsewhere). 
 Additional parameters are $\Omega_m$ and $H_0$. \\
\noindent   
 3) Use flux-averaging statistics in 
computing $\chi^2$ of SN Ia data \cite{Wang00b}.\\
\noindent
 4) Add other data (CMB, LSS) to help constrain $\Omega_m$.\cite{Knop}\\
\noindent 
  5) Use Markov Chain Monte Carlo (MCMC) to find marginalized
probability distributions of estimated parameters. 

I will highlight the need for flux-averaging of supernovae and the advantage
of MCMC below.

\subsection{Minimizing gravitational lensing effect on SNe Ia by flux-averaging}

Since our universe is inhomogeneous in matter distribution, the apparent brightness
of SNe Ia can be altered by the intervening matter distribution.
\cite{Wang02} gives a universal probability distribution for weak lensing 
magnification of SNe Ia.
Because lensing only redistributes the light (flux) from the SNe Ia,
the effect of weak lensing on supernovae can be minimized by flux-averaging.
Therefore, flux-averaging justifies the use of 
the distance-redshift relation for a smooth universe in the analysis
of SN Ia data. \cite{Wang00b}.

For $\chi^2$ statistics
using MCMC or a grid of parameters, 
here are the steps in flux-averaging:

(1) Convert the distance modulus of SNe Ia, $\mu_0(z)$,  into 
``fluxes''
\be
F(z_j) \equiv 10^{-(\mu_0(z_j)-25)/2.5} =  
\left( \frac{d_L^{data}(z)} {\mbox{Mpc}} \right)^{-2}.
\ee

(2) For a given set of cosmological parameters $\{ {\bf s} \}$,
obtain ``absolute luminosities'', \{${\cal L}(z_j)$\}, by
removing the redshift dependence of the ``fluxes'', i.e.,
\be
{\cal L}(z_j) \equiv d_L^2(z_j |{\bf s})\,F(z_j).
\ee

(3) Flux-average the ``absolute luminosities'' \{${\cal L}^i_j$\} 
in each redshift bin $i$ to obtain $\left\{\overline{\cal L}^i\right\}$:
\be 
 \overline{\cal L}^i = \frac{1}{N}
 \sum_{j=1}^{N} {\cal L}^i_j(z^i_j),
 \hskip 1cm
 \overline{z_i} = \frac{1}{N}
 \sum_{j=1}^{N} z^i_j. 
\ee

(4) Place $\overline{\cal L}^i$ at the mean redshift $\overline{z}_i$ of
the $i$-th redshift bin, now the binned flux is
\be
\overline{F}(\overline{z}_i) = \overline{\cal L}^i /
d_L^2(\overline{z}_i|\mbox{\bf s}).
\ee
The 1-$\sigma$ error on each binned data point $\overline{F}^i$,
$\sigma^F_i$, is taken to be the root mean square of the 1-$\sigma$ errors on the 
unbinned data points in the $i$-th redshift bin, \{$F_j^i$\} ($j=1,2,...,N$), 
multiplied by $1/\sqrt{N}$ (see \cite{Wang00a}).

(5) Apply ``flux statistics'' to the flux-averaged data, 
$\left\{\overline{F}(\overline{z}_i)\right\}$:
\be
\label{eq:chi2}
\chi^2 (\mbox{\bf s}) = \sum_i \frac{ \left[\overline{F}(\overline{z}_i) -
F^p(\overline{z}_i|\mbox{\bf s}) \right]^2}{\sigma_{F,i}^2},
\ee
where $F^p(\overline{z}_i|\mbox{\bf s}_1)=
\left( d_L(z|\mbox{\bf s}) /\mbox{Mpc} \right)^{-2}$.

A Fortran code that uses flux-averaging statistics 
to compute the likelihood of an arbitrary 
dark energy model (given the SN Ia data from \cite{Riess04})  
can be found at $http://www.nhn.ou.edu/\sim wang/SNcode/$
\cite{Wang00b,WangMukherjee,WangTegmark}.

\subsection{Likelihood analysis using Markov Chain Monte Carlo (MCMC)}

In a likelihood analysis, we need to find the high confidence
region of our parameter space given the data.
Traditionally, this is done by using a grid of parameter values
that span the allowed parameter space.

Markov Chain Monte Carlo (MCMC) uses the 
Metropolis algorithm: \\
\noindent
(1) Choose a candidate set of parameters, 
${\bf s}_*$, at random from a proposal distribution. \\
\noindent
(2) Accept the candidate set of parameters with probability $A({\bf s}, {\bf s}_*)$; 
otherwise, reject it. 

For Gaussian distributed observables, the acceptance function
\be
A({\bf s}, {\bf s}_*) = \min\left\{ 1, \exp\left[-\chi^2({\bf s}_*)+\chi^2({\bf s})
\right]\right\}.
\ee
For sufficient sampling, the true probability density functions (pdf's) are recovered. 
See \cite{Neil} for a review of MCMC.

MCMC is much faster than the grid method, 
and scales linearly with the number of parameters. 
It gives smooth probability density functions (pdf's) for estimated parameters 
since they receive contribution from all MCMC samples (each with 
a random set of parameter values). 
Public software containing MCMC is available \cite{LB02}.

\section{Status and Outlook}

Using the latest SN Ia data, the Riess sample \cite{Riess04}, 
which contains 6 SNe Ia at $z>1.25$, together with CMB and LSS data,
\cite{WangTegmark} obtained the most accurate measurements to date of the 
dark energy density $\rho_X$ as a free function of cosmic time (see 
left panel of Fig.3).
The solid lines indicate the 68\% (shaded) and 95\% confidence regions 
of $\rho_X(z)$ parametrized by its values at $z=0.467, 0.933, 1.4$ 
(splined elsewhere for $z<1.4$).
The thick and thin dashed lines indicate the estimated $\rho_X(z)$
parametrized by its values at $z=0.7, 1.4$ 
(splined elsewhere for $z<1.4$).
A powerlaw form of $\rho_X(z)$ is assumed at $z>1.4$ (where 
there are only two observed SNe Ia); this allows the CMB and LSS constraints to
be included in a self-consistent manner.
This is important since it is incorrect (although common practice) to
include priors on $\Omega_m$ derived assuming $\rho_X(z)/\rho_X(0)=1$. \cite{WangTegmark}
Further, $\rho_X(z)$ derived assuming fixed values of $\Omega_m$ may be misleading
because of the degeneracy between $\rho_X(z)$ and $\Omega_m$.
All our results have been marginalized over $\Omega_m$ and $H_0$.

\begin{figure}[!hbt]
\begin{center} 
\psfig{file=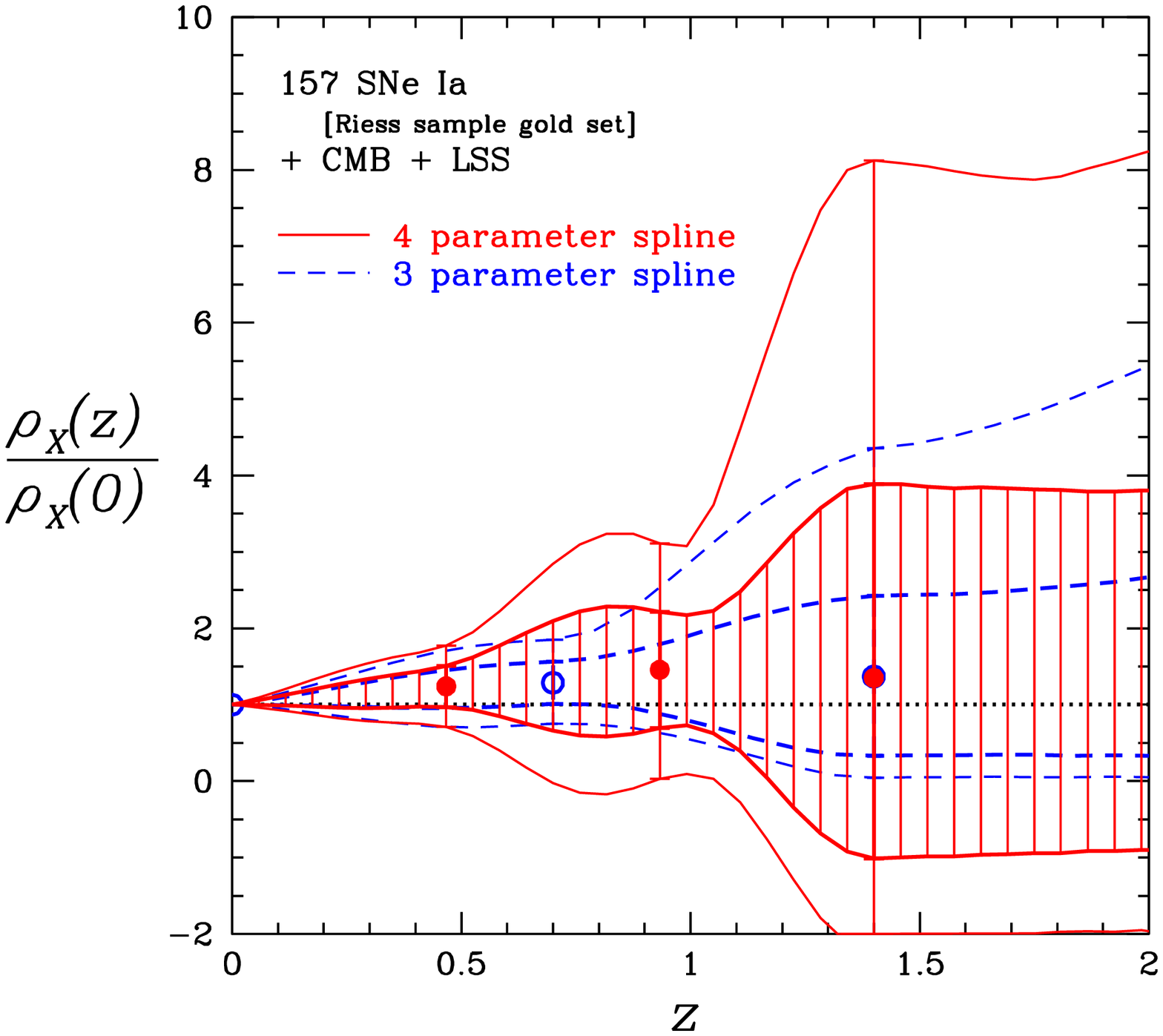,width=6.8cm} 
\psfig{file=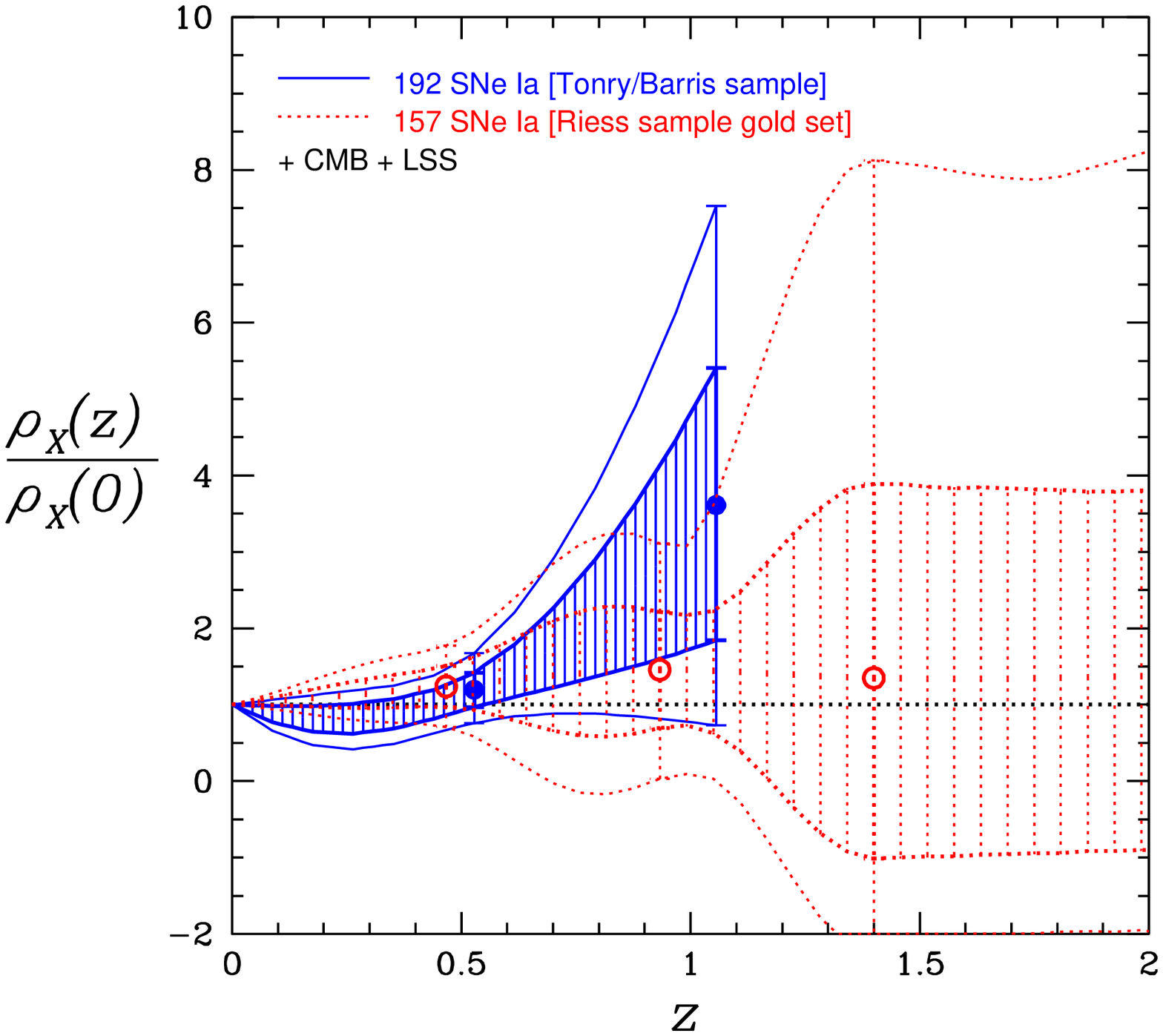,width=6.8cm} 
\vskip -0.6cm
\caption{\footnotesize%
Dark energy density $\rho_X$ as a free function of cosmic time
estimated from supernova, CMB (shift parameter 
$R_0 = 1.716 \pm 0.062$), and LSS (growth parameter $f_0 \equiv 
f(z=0.15) =0.51\pm 0.11$) data.
Left panel: The latest SN data, with $\rho_X$ parametrized by 4 and 3 parameter splines;
with estimated $\Omega_m=0.278 \pm  0.061$ and $\Omega_m=0.279  \pm 0.053$ respectively. 
Right panel: The latest SN data (Riess sample) compared to previous data (Tonry/Barris sample,
with estimated $\Omega_m=0.332 \pm 0.062$).
} 
\end{center} 
\end{figure}

We compare the $\rho_X(z)$ estimated from the
latest SN data (Riess sample) to that from the previous data 
(Tonry/Barris sample) in the right panel of Fig.3. 
Note that the Riess sample gives a substantially tighter constraint
on $\rho_X(z)$ than the Tonry/Barris sample, and somewhat smaller $\Omega_m$. 
Although the two estimates overlap at the 68\% confidence level,
the Riess sample clearly favors a dark energy density less
dependent on time (with the cosmological constant model giving
an excellent fit to the data).

It is interesting that current data already rule out dark energy models
with $\rho_X(z)$ that vary a lot with time. Future deep supernova surveys 
on a dedicated telescope will greatly increase the number of observed supernovae
\cite{Wang00a}. This will allow improved calibration of SNe Ia
as cosmological candles, leading to precise measurement of
the dark energy density as function of cosmic time, which
will have historical impact on particle physics and cosmology.

\begin{ack}    
I would like to thank David Cline for an interesting and productive meeting,
and Katie Freese, Peter Garnavich, Pia Mukherjee, Greg Tarle, Max Tegmark, and
Paul Steinhardt for helpful discussions.
This work was supported in part by
NSF CAREER grant AST-0094335.     
\end{ack}

\end{document}